\documentclass[aps,prd,showpacs,nofootinbib,floats,floatfix,preprintnumbers,groupedaddress,twocolumn]{revtex4-1}
\usepackage{graphicx,epsfig}
\usepackage{dcolumn}
\usepackage{bm}
\usepackage{latexsym}
\usepackage{color}
\usepackage{tabularx}
\usepackage{ulem}
\usepackage{hyperref}
\usepackage{float}
\usepackage{tabularx}
\usepackage{color}
\usepackage{hyperref}
\usepackage{colortbl}
\usepackage{listings}
\usepackage{amsmath,amsfonts,amssymb}
\usepackage{fancyhdr}
\usepackage{hyperref}
\usepackage{natbib}
\hypersetup{
	colorlinks   = true, 
	urlcolor     = blue, 
	linkcolor    = blue, 
	citecolor   = red 
}

\hypersetup{
	colorlinks   = true, 
	urlcolor     = blue, 
	linkcolor    = blue, 
	citecolor   = red 
}

\begin{document}
\title{Presence of horizon makes particle motion chaotic}
\author{Surojit Dalui}
\email{suroj176121013@iitg.ac.in}
\author{Bibhas Ranjan Majhi}
\email{bibhas.majhi@iitg.ac.in}
\author{Pankaj Mishra }
\email{pankaj.mishra@iitg.ac.in}
\affiliation{Department of Physics, Indian Institute of Technology Guwahati, Guwahati 781039, Assam, India}

\date{\today}

\begin{abstract}
We analyze the motion of a {\it massless} and {\it chargeless} particle very near to the event horizon. It reveals that the radial motion has exponential growing nature which indicates that there is a possibility of inducing chaos in the particle motion of an integrable system when it comes under the influence of the horizon. This is being confirmed by investigating the Poincar$\acute{e}$ section of the trajectories with the introduction of a harmonic trap to confine the particle's motion. Two situations are investigated: (a) {\it any} static, spherically symmetric black hole and, (b) spacetime represents a stationary, axisymmetric black hole (e.g., Kerr metric). In both cases, the largest Lyapunov exponent has upper bound which is the surface gravity of the horizon. We find that the inclusion of rotation in the spacetime introduces more chaotic fluctuations in the system. The possible implications are finally discussed.
\end{abstract}

\pacs{04.62.+v,
04.60.-m}
\maketitle

{\section{\label{Intro}Introduction}}
The dynamics of a particle around a massive object is one of the central attention in physics. General theory of relativity (GR) successfully explains several phenomena in the motion of astrophysical objects. It explains how the light can bend due to the presence of a mass in the spacetime. Among several compact objects, black holes are one of the fascinating thing in the Universe till now. Theoretically these are the solutions of Einstein's equations of motion and defined by the region whose boundary is a one way membrane (classically through which nothing can escape), known as event horizon. Recent discovery of LIGO \cite{Abbott:2016blz} confirms that black holes are no longer a theoretical concept, rather they indeed have the  existence in the spacetime. Till today, researchers are devoting lot of attentions not only to understand the Physics of black holes, but also the kind of phenomena that they induce around themselves both at the astrophysical and the quantum level.

Near horizon Physics, at the classical as well as quantum levels, is very important in various ways. Recently, a quantum mechanical treatment explores that fact that the presence of Killing horizon makes the particle motion Brownian when seen from an accelerated frame \cite{Adhikari:2017gyb}.  In this paper, we try to understand how the motion of a particle changes when it approaches very near to the black hole horizon (which is an event horizon). This is investigated completely at the classical way. There have been studies in this direction \cite{Bombelli:1991eg,Sota:1995ms,Vieira:1996zf,  Suzuki:1996gm,Cornish:1996ri,deMoura:1999wf,Hartl:2002ig,Han:2008zzf,Takahashi:2008zh,Hashimoto:2016dfz,Li:2018wtz} {\it in presence of perturbation}, either externally or internally due to change of any black hole parameter. These calculations are performed mainly based on either Newtonian approximation \cite{Sota:1995ms,Vieira:1996zf} or  effective potential technique \cite{Bombelli:1991eg,Suzuki:1996gm,Cornish:1996ri,deMoura:1999wf,Takahashi:2008zh}. Moreover, black hole system is either spinning \cite{Takahashi:2008zh} or magnetized \cite{Li:2018wtz} with the massive, charged or spinning test particles \cite{Hartl:2002ig,Han:2008zzf}. There is a recent analysis \cite{Hashimoto:2016dfz} that considers the effect of Schwarzschild black hole on a massive test particle in presence of harmonic potentials as perturbation. To make sure that the particle does not enter into the horizon an extra potential has been added to the system. It has been observed in all cases that the motion of the particle is chaotic in nature. However, {\it the situation with the massless particles have not been addressed so far}. There has been a sporadic attempt to understand the behaviour of Lyapunov exponent for the case of null trajectories of a particle \cite{Cardoso:2008bp}, but this does not give a complete understanding of the situation. Till now, this has been a thriving area of the research since last few years as we  know that the interaction of light with gravity is quite non-trivial in nature.  

In this paper we have addressed the issue more deeply and in a more general way. Calculations show that the outgoing radial trajectory of the particle only on the black hole spacetime, when it is very near to the horizon, grows exponentially with time. It indicates that there is a {\it possibility of induction of chaotic nature} in the motion of the particle, which is already under a integrable system, when it comes to the influence of the horizon. We ascertain this observation numerically using detailed investigation of  the Poincare section of the particle trajectories, which is under harmonic potentials, comes to the influence of the black hole event horizon. Both the situations, {\it any static spherically symmetric} (SSS) (not necessarily Schwarzschild case) and {\it stationary axisymmetric} (Kerr solution) black holes, have been investigated. The test particle has been considered to be {\it chargeless} and {\it massless}. Here we find that the particle trapped under harmonic potential can be considered as an integrable system, however, the nature of the particle trajectories changes as the system approaches to the black hole horizon. It is observed that a static spherically symmetric horizon induces chaos in a particular energy range, while, the presence of rotation in the spacetime makes the system more chaotic.

In this analysis we have not only found that the chaos is inevitable in presence of horizon, the value of Lyapunov exponent has also an upper bound. Interestingly, the bound can be shown to be given by $\lambda_L\leq\kappa$, where $\kappa$ is the surface gravity of the black hole (it is the acceleration of a particle, measured by an asymptotic distance observer, which is very near to the horizon). This has been predicted recently \cite{Hashimoto:2016dfz} for a massive particle in a completely different analysis. We found that such a value is compatible with the bound predicted in \cite{Maldacena:2015waa} by analyzing the out-of-order correlator of some observables in the Sachdev-Ye-Kitaev (SYK) model. This shows that the bound is very much universal in nature.

Let us now discuss what we can predict by our present analysis. Note that we consider only the massless particles which are following the outgoing trajectories, of which the radial one is the radial null geodesic. It is observed that the same null geodesic is responsible for the Hawking radiation \cite{Hawking:1974rv,Hawking:1974sw} of the particles from the horizon (see \cite{Parikh:1999mf,Banerjee:2008gc,Banerjee:2008ry} for understanding the Hawking radiation as tunneling). Therefore, it may be worth mentioning that the radiated particles, after escaping from the horizon barrier, exhibit chaotic behavior in their motion due to the influence of the horizon as well as other external perturbation due to presence of various objects in the Universe. This implies that {\it the horizon not only radiates (Hawking radiation), it also infuses chaos}. Of course, this is not a conclusive statement, rather a suggestive one. In order to get more insight to this one needs to investigate more, may be in the quantum mechanical way. After having this discussions,  we now turn our focus to the main analysis.
\section{\label{Sphere}Static spherically symmetric black hole}
Consider a static, spherically symmetric black hole background, given by
\begin{equation}
ds^2=-f(r)dt^2+\frac{dr^2}{f(r)}+r^2 d\Omega^2~,
\label{1.01}
\end{equation}
where the horizon $r=r_H$ is determined by $f(r=r_H)=0$ and $d\Omega^2= (d\theta^2+\sin^2\theta d\phi^2)$. It has a coordinate singularity at this position. To remove this let us adopt the Painleve coordinate transformation:
\begin{equation}
dt\rightarrow dt-\frac{\sqrt{1-f(r)}}{f(r)}dr~.
\label{1.02}
\end{equation}
Under this transformation, the above metric takes the following form:
\begin{equation}
ds^2=-f(r)dt^2+2\sqrt{1-f(r)}dtdr+dr^2+r^2d\Omega^2~.
\label{1.03}
\end{equation}
It has a timelike Killing vector $\chi^a=(1,0,0,0)$ and the energy of a particle, moving under this background, is given by $E=-\chi^ap_a=-p_t$, where $p_a=(p_t,p_r,p_\theta,p_\phi)$ is the four momentum vector. Our aim in the following is to find this energy in terms of other components of momentum. To find it, we take help of the covariant form of the dispersion relation $g^{ab}p_ap_b=-m^2$, with $m$, mass of the particle. Expanding this under the background (\ref{1.03}), we obtain
\begin{equation}
E^2+2\sqrt{1-f(r)}~p_rE-\Big(f(r)p_r^2+\frac{p_\theta^2}{r^2}\Big)=m^2~,
\label{1.04}
\end{equation}
where, only the radial and $\theta$ directions motion have been considered, i.e., the particle is moving only along the radial and the $\theta$ directions. For a massless particle, solution of the above equation (\ref{1.04}) with $m=0$ gives the energy. It is found that it has two values:
\begin{equation}
E=-\sqrt{1-f(r)}~p_r\pm\sqrt{p_r^2+\frac{p_\theta^2}{r^2}}~.
\label{1.05}
\end{equation}
Positive sign denotes the energy for the outgoing particle, while, the other sign is for ingoing particle. In this paper we will be mainly interested to investigate the dynamics of the outgoing particles, therefore, throughout the discussions only the positive sign will be considered.

Next aim is to find the trajectory of the particle. It will be computed from the Hamilton's equations of motion.  Before that concentrate for the moment only on the {\it very near horizon radial motion}. The equation of it (taking $p_\theta=0$) for the energy (\ref{1.05}) is 
$\dot{r}=\partial E/\partial p_r = -\sqrt{1-f(r)}+1$.
Make an expansion of  $f(r)$ near the horizon as
\begin{equation}
f(r) \simeq 2\kappa(r-r_H)~,
\label{1.10}
\end{equation}
where, we have retained only the first order term. Here $\kappa = f'(r_H)/2$ is the surface gravity of the black hole. Substitution of this in $\dot{r}$ equation leads to 
\begin{equation}
\dot{r}\simeq \kappa(r-r_H)~.
\label{1.11}
\end{equation}
The solution is
$r=r_H+Cr_He^{\kappa\lambda}$,
where, $C$ is the integration constant and $\lambda$ is parameter with respect to which the derivative is taken, i.e., $\bf{.}\equiv \frac{d}{d\lambda}$. The above solution implies the exponential growth of radial coordinate which can be attributed to the appearance of chaos in an integrable system when it comes to the influence of the horizon. This can be expected as in GR even a photon can have bound circular motion which is not expected in Newtonian approximation \cite{Carroll}. Note that the actual trajectories which will be evaluated and investigated later part of the paper are highly non-linear in four-dimensional phase space. We shall show explicitly that this indeed the case by keeping this particle in a harmonic potential and allowing it to move along the $\theta$ direction as well. In this situation, it may be worth to point out that the Lyapunov coefficient ($\lambda_L$) is bounded as \cite{comment1}:
\begin{equation}
\lambda_L \leq \kappa~.
\label{1.13}
\end{equation}
Now if one consider the quantum nature of the black hole, then it has a temperature, given by the Hawking expression $T=\hbar\kappa/2\pi$. In that case the above bound reduces to a very well known form $\lambda_L\leq 2\pi T/\hbar$. This was first mentioned in \cite{Maldacena:2015waa} for the SYK model.

There is a very interesting connection with the above radial null geodesic (\ref{1.11}) with the Hawking effect in the context of tunneling mechanism \cite{Parikh:1999mf,Banerjee:2008gc,Banerjee:2008ry}. It must be noted that precisely this path has been used to find the tunneling probability from the horizon for the outgoing particles. One finds that this is non-zero and leads to Hawking radiation with the temperature, mentioned above. Therefore, after escaping from the horizon the radiated particles may exhibit chaotic motion due to the influence of the horizon as well as the perturbation induced by the presence of other objects in the Universe. Hence the current analysis implies that the horizon not only radiates, it also makes the radiation chaotic.

With this let us now confirm if the presence of horizon really creates chaos in the system. For that we consider two harmonic potentials $(1/2)K_r(r-r_c)^2$ and $(1/2)K_\theta(y-y_c)^2$ along $r$ and $\theta$ directions, respectively \cite{comment2}. Here $y=r_H\theta$, $K_r$ and $K_\theta$ are spring constants while $r_c$ and $y_c$ are the equilibrium positions of these two harmonic potentials. The massless particle is initially under these potentials. Now this system is kept in the influence of the black hole horizon as well. Our aim is to investigate the collective impact of horizon on this integrable system, particularly the nature of the particle trajectories. Here we choose the form of black hole metric $f(r)$ as (\ref{1.10}). Now if the particle moves under the influence of these potentials under the background (\ref{1.03}) then the total energy of this particle is
\begin{eqnarray}
E=&&-\sqrt{1-f(r)}~p_r+\sqrt{p_r^2+\frac{p_\theta^2}{r^2}}+\frac{1}{2}K_r(r-r_c)^2
\nonumber
\\
&&+\frac{1}{2}K_\theta(y-y_c)^2~.
\label{1.14}
\end{eqnarray}
and correspondingly, the equations of motion will have the form as 
\begin{eqnarray}
&&\dot{r}=\frac{\partial E}{\partial p_r} = -\sqrt{1-f(r)}+\frac{p_r}{\sqrt{p_r^2+\frac{p_\theta^2}{r^2}}}~;
\label{1.15}
\\
&&\dot{p_r} = -\frac{\partial E}{\partial r} = -\frac{f'(r)}{2\sqrt{1-f(r)}}p_r+\frac{p_\theta^2/r^3}{\sqrt{p_r^2+p_\theta^2/r^2}}
\nonumber
\\
&&- K_r(r-r_c)~;
\label{1.16}
\\
&&\dot{\theta} = \frac{\partial E}{\partial p_\theta} =\frac{p_\theta/r^2}{\sqrt{p_r^2+p_\theta^2/r^2}}~;
\label{1.17}
\\
&&\dot{p_\theta} = -\frac{\partial E}{\partial\theta}=-K_\theta r_H(y-y_c)~.
\label{1.18}
\end{eqnarray}
In the above the interaction between the harmonic potentials and the black hole spacetime has been taken as very weak so that this can be ingnored compared to the other terms in (\ref{1.14}). 
These are the main equations which we will use to study the motion of the particle numerically.

\section{\label{Kerr}Kerr black hole}
After having the discussion on the effect of the SSS black hole on the particle in this section we investigate the effect of rotation of the black hole on the over all dynamics the system. For this we can proceed in the similar way like we did for SSS. The Kerr metric in dragging Painleve coordinate is given in \cite{Jiang:2005ba}.
Due to the large size of the equations, the metric and the trajectories for this case have been provided in Appendix \ref{App}. It may be noted that here also if one concentrates only on the radial trajectories, it is given by (setting $\theta=0$ and $p_{\theta}=0$ in eq. (\ref{1.33}) of Appendix \ref{App})
\begin{eqnarray}
&&\dot{r}=1-\sqrt{\frac{a^2r+2r_H^3-rr_H^2}{r_H(a^2+r_H^2)}}
\nonumber
\\
&&=1-\sqrt{1-2\kappa(r-r_H)}\simeq \kappa(r-r_H)~,
\end{eqnarray}
where,
\begin{equation}\label{kerrK}
\kappa=(r_H^2-a^2)/(2r_H(r_H^2+a^2)).
\end{equation}
Note that in this case also we are getting the same radial equation in the near horizon limit. So again the same solution. As a result of this the Lyapunov coefficient bound  would be given by (\ref{1.13}). Therefore we can see that the rotating black hole also appears to infuse chaotic fluctuations in the motion of particles. In the next section we will ascertain this observation by numerically solving the full set of dynamical equations and will also investigate the detailed influence of the rotation on the chaotic dynamics of the particles.

Before going to the numerical analysis, it is very useful to check if any other external potential (like due to the back reaction of the different fields on the spacetime) can effect the bound (\ref{1.13}) on the Lyapunov exponent {\footnote{We thank the anonymous referee for pointing out this topic.}}.
For that suppose we consider an arbitrary potential which is function of radial coordinate only, i.e. $V(r)$. This assumption is very much convenient for more realistic potentials, made by some other fields condensation with a back reaction to the gravity metric, one usually encounters. As for example, in case of scalar fields the potential near to the horizon is of the form $\ln (x/C_0)$, while, for electrostatic case it is proportional to $x$ \cite{Hashimoto:2016dfz}, where $x=r-r_H$ and $C_0$ is a positive constant. Under these circumstances, the equations of motions for SSS case will be given by (\ref{1.15}) -- (\ref{1.18}) with harmonic term in (\ref{1.16}) will be replaced by ($- dV/dr$). With this we find that there will be no contribution of the potential term in the Eq. (\ref{1.18}). 
Now to understand the behaviour of radial motion, as earlier, we need to concentrate only on the radial equation by setting $p_\theta=0$. Since we are using the massless condition, it can be noted that although the equation corresponding to $\dot{p_{r}}$ will be affected by the term like $dV(r)/ dr$, the equation of $\dot{r}$ (Eq. (\ref{1.15})) will not get affected as it is independent of $p_{r}$.  Hence the motion along the radial direction will remain unaffected by introducing any arbitrary potential of the form $V(r)$. Therefore the solution of $\dot{r}$ will always be given by that of Eq. (\ref{1.11}) and correspondingly the Lyapunov exponent is bounded by the surface gravity $\kappa$. However for the massive particle mass factor will appear in the equation of $\dot{r}$ and the form of $V(r)$ will affect the motion in the radial direction as it will not be independent of $p_{r}$.  As a result, the solution of $\dot{r}$ will get changed and then this change will certainly appear in the bound of $\lambda$. This particular features have  been explicitly discussed in \cite{Hashimoto:2016dfz}. Therefore overall we find that the addition of any radial potential will not affect the upper bound of the Lyapunov coefficient for massless particle. Arguing on the similar line one can check that addition of the above mentioned potential $V(r)$ will also not affect the bound of Lyapunov exponent in the case of Kerr Black Hole (See equations (\ref{1.33}) -- (\ref{1.36})). 
\section{\label{numeric}Numerical Analysis}
In the last section we analytically showed that the presence of black hole horizon induces the exponential growth of the radial trajectory of the particles which indicated the chaotic behaviour for both SSS and Kerr black holes. In this section we ascertain the claim by analyzing the Poincar$\acute{e}$ section of the system obtained by solving the dynamical equations of motion for  SSS (Eqs.~\ref{1.15}-\ref{1.18}) and Kerr (see Eqs. (\ref{1.33})--(\ref{1.36}) of Appendix \ref{App}) black holes. First we present numerical results for the SSS black hole and then those for the Kerr black hole in which we systematically analyze the effects of the rotation parameter on the chaotic fluctuations.  

\begin{figure}[!ht]
\centering
 \includegraphics[scale=0.20, angle=0]{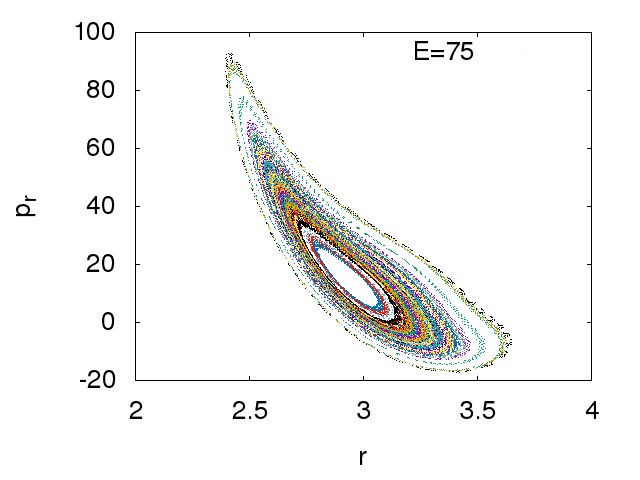}
 \includegraphics[scale=0.20, angle=0]{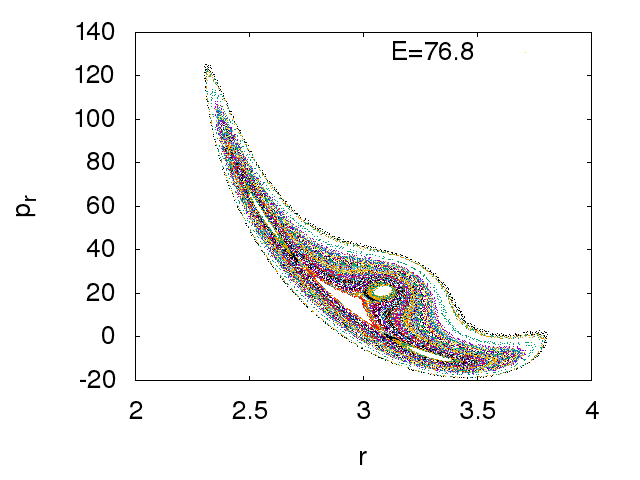}
 \includegraphics[scale=0.20, angle=0]{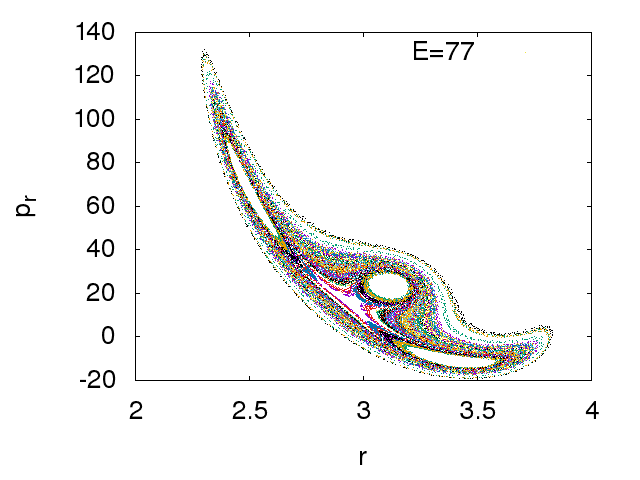}
 \includegraphics[scale=0.20, angle=0]{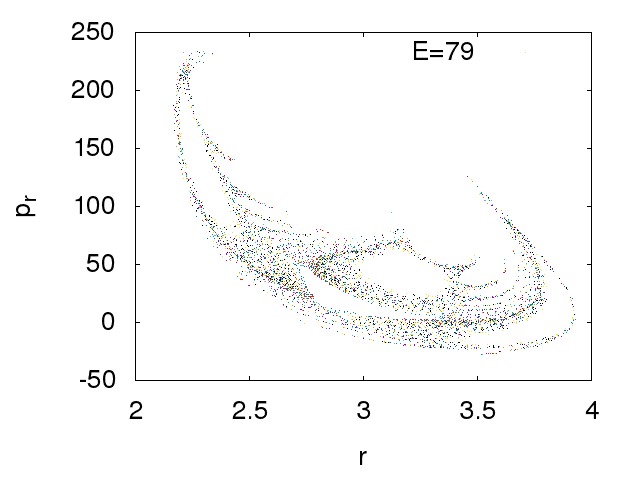}
 \caption{(Color online) The Poincar$\acute{e}$ sections in the ($r$, $p_r$) plane with $\theta=0$ and $p_{\theta}>0$ at different energies for the SSS black hole. The energies are  $E=75$,  $E= 76.8$,  $E= 77$, and $E= 79$. The other parameters are $r_H=2.0$, $\kappa=0.25$, $r_c=3.2$, $\theta_c=0$, $K_r=100$ and $K_{\theta}=25$.  For large energy the KAM Tori break and the entire region gets filled with the scattered points indicating the presence of chaos. }
 \label{fig:schwar}
\end{figure}

\begin{figure}[!ht]
 \centering
 \includegraphics[scale=0.20, angle=0]{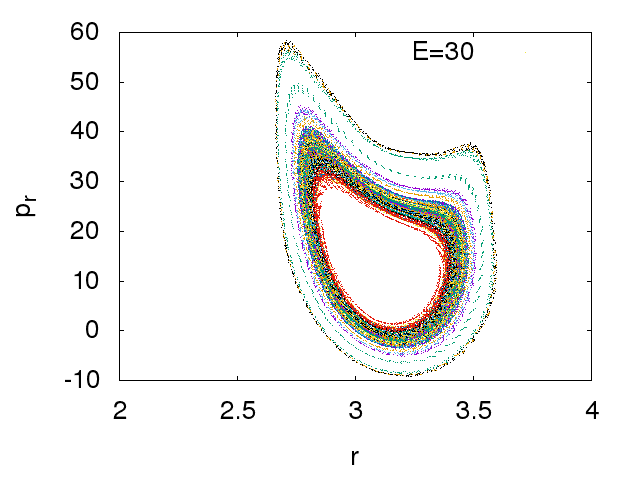}
 \includegraphics[scale=0.20, angle=0]{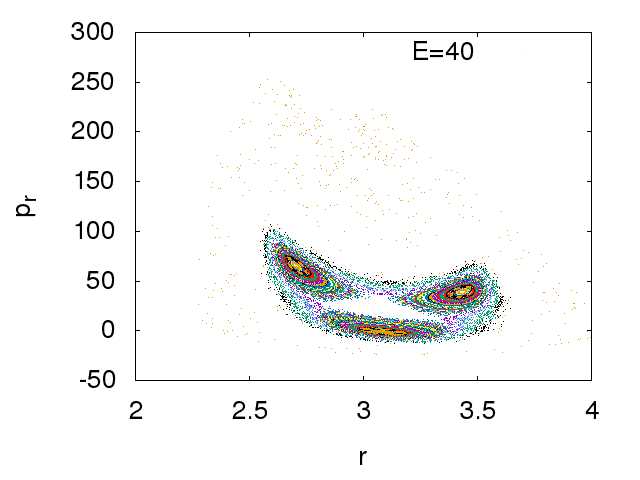}
 \includegraphics[scale=0.20, angle=0]{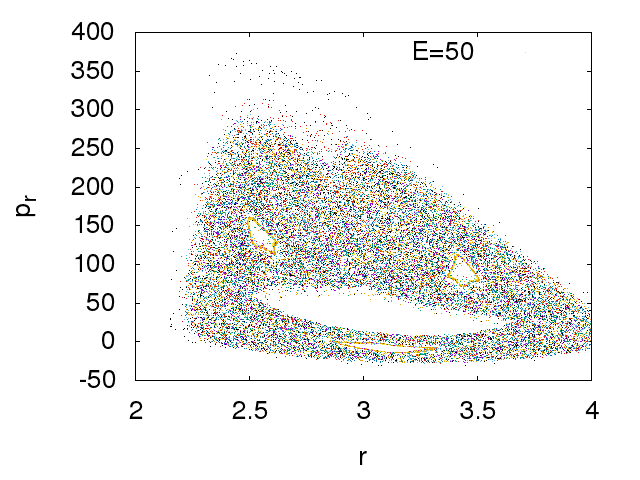}
 \includegraphics[scale=0.20, angle=0]{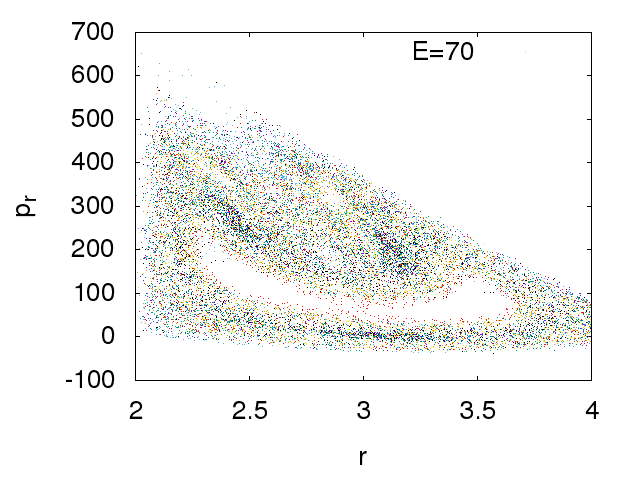}
 \caption{(Color online) The Poincar$\acute{e}$ sections in the ($r$, $p_r$) plane with $\theta=0$ and $p_{\theta}>0$ for different energy for the Kerr black hole model at fixed rotation parameter $a=0.9$. The energies are  $E=30$,  $E= 40$,  $E= 50$, and $E= 70$. The other parameters are same as in Fig.~\ref{fig:schwar}.  For large energy the KAM Tori break and the entire region is filled with the scattered points indicating the chaotic trajectory of the particles. }
 \label{fig:kerrap9}
\end{figure}
For SSS black hole model, the dynamical equations (Eqs.~\ref{1.15}-\ref{1.18}) are numerically solved using the fourth order Runge-Kutta scheme with fixed $dt=10^{-3}$. For present study we have considered $K_r=100$, $K_{\theta}=25$, $r_H=2$, $\kappa=0.25$, $r_c=3.2$ and $\theta_c=0$. The variables  $r$, $\theta$, and $p_{r}$ are initialized with the random numbers and $p_{\theta}$ is obtained from Eq.~\ref{1.05} for a fixed energy $E$. 

In Fig.~\ref{fig:schwar} we show the Poincar$\acute{e}$ section of the particle trajectory (for SSS type black hole) projected over the ($r$,$p_r$) plane for different energies.  The section is defined by the condition $p_{\theta}>0$ and $\theta=0$.  We have considered the energies $E=75, 76.8, 77$, and $79$ as indicated in the plots. For low energy $E=75$ the Poincar$\acute{e}$ section exhibits the regular KAM (Kolmogorov-Arnold-Moser) tori~\cite{nonlinear:02} and the corresponding orbit is mainly confined near the center of the harmonic potential ($r_c=3.2$). Different colors in the figures indicate the trajectory of the particles for different initial conditions. As the total energy of the system is increased the trajectory approaches near to the black hole horizon ($r_H=2$) as a consequence of this KAM tori starts getting distorted and appeared to be pinched as shown in the figure for E=76.8 and E=77. Further increase in the energy ($E=79$) results the complete breaking of regular Tori and appearance of scattered points in the plane. This feature of the Poincar$\acute{e}$ section supports the chaotic nature of the particle trajectory near the horizon as showed in the last section. Due to presence of horizon we have some upper bound on the energy. At present increasing the energy further above $E=79$ we find that during time evolution $r$ becomes less than the position of the horizon ($r_H=2$) that brings numerical instability in our calculation.   

\begin{figure}[!ht]
\centering
 \includegraphics[scale=0.20, angle=0]{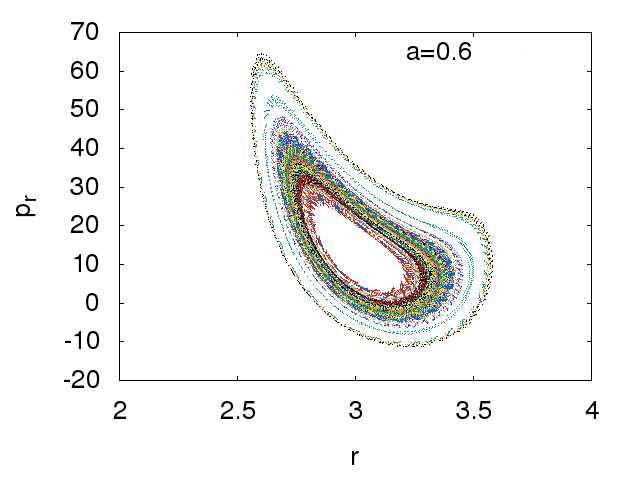}
 \includegraphics[scale=0.20, angle=0]{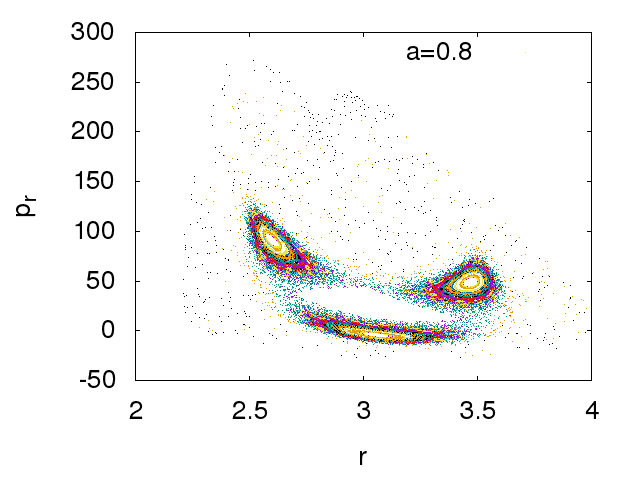}
 \includegraphics[scale=0.20, angle=0]{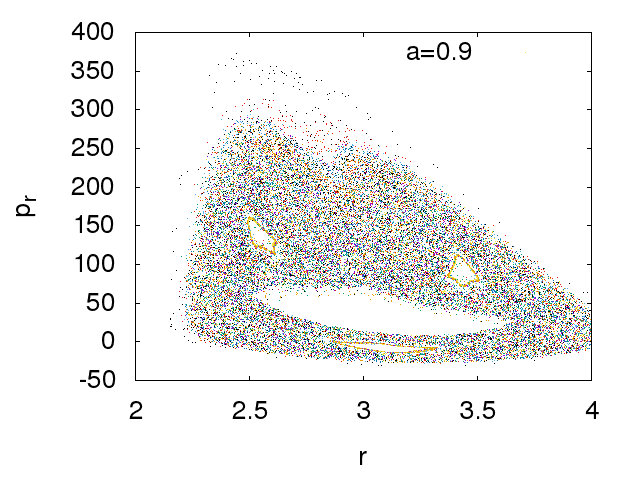}
 \includegraphics[scale=0.20, angle=0]{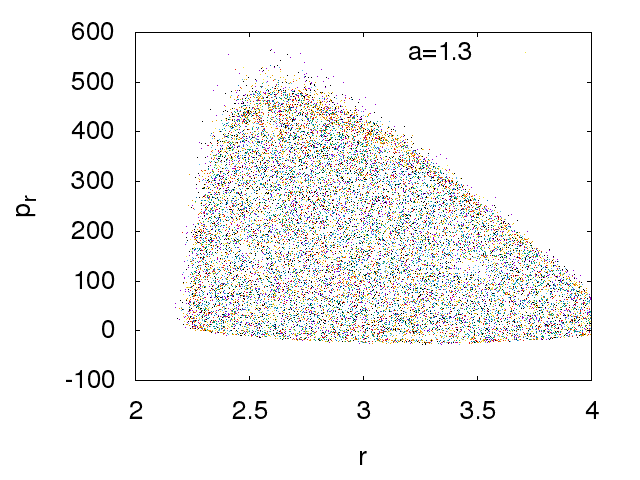}
 \caption{(Color online) The Poincar$\acute{e}$ sections in the ($r$, $p_r$) plane with $\theta=0$ and $p_{\theta}>0$ for fixed energy E=50 with different rotation parameter $a$. The rotation parameter  $a=0.6$,  $a= 0.8$,  $a=0.9 $, and $a= 1.3$, respectively from the top to the bottom. The other parameters are $r_H=2.0$, and $K_r=100$ and $K_{\theta}=25$. Increase in the rotation induces the chaos in the dynamics of the particles. }
 \label{fig:kerrE50}
\end{figure}

In order to see the effect of the rotation of the black hole on the particle dynamics next we consider the Kerr black hole model. The corresponding dynamical equations (see Eqs. (\ref{1.33})--(\ref{1.36}) of Appendix \ref{App}) are solved using the Runge-Kutta fourth order. The initial conditions are chosen in the similar line as discussed for the SSS model.  In Fig.~\ref{fig:kerrap9} we show the Poincar$\acute{e}$ sections for different energies with fixed rotation parameter $a=0.9$. All the other parameters are considered same as the SSS model. We observe the similar feature of the KAM tori upon increase of energy for Kerr black hole as we observed for the SSS model. The regular KAM tori appeared at low energy $E=30$ gets squeezed along with appearance of some region filled with the scattered points as the energy is increased to $E=40$. At high energy ($E=50,70$) as the surface of the trajectory approaches near the horizon there is complete breaking of these tori which is quite evident with the filling of the region with the random points. Interestingly here we obtain that the chaotic nature appears at relatively lower energy than those obtained with the SSS black hole. This particular feature suggests that the rotation of the black hole may introduce more chaotic fluctuations in the trajectories of the particles approaching towards the horizon. We confirm this by analyzing the nature of the particle trajectory by changing the rotation parameter $a$ for fixed energy $E=50$. In Fig.~\ref{fig:kerrE50} we plot the Poincar$\acute{e}$ section in the plane $(r,p_r)$ for different rotation parameters $a=0.6, 0.8, 0.9$, and $1.3$ at fixed energy $E=50$. We clearly find that the increase in $a$ introduces chaotic fluctuations in the trajectories and at very high rotation $a=0.9 1.3$ the trajectory becomes fully chaotic. The nature of the appearance of chaos in the system upon increase in the $a$ for fixed energy appears to be same as those obtained while increasing the energy for fixed $a=0.9$ (see Fig.~\ref{fig:kerrap9}).

So far our numerical calculation is based on the near horizon approximated metric. This has been done as we are interested to investigate the influence of horizon on a particle dynamics when it is very near to the horizon. In this region we must have $r-r_H<r_H$. It may be noted that the condition is well incorporated in the simulations as all the data points in the figures are confined within such range. In the similar context, the same type of approximation has also been taken recently in \cite{Hashimoto:2016dfz}. Here we emphasise that the use of near horizon approximation has nothing to do with the appearance of chaotic behaviour in the system. To confirm this fact, we also perform the same numerical analysis without any approximation in the metric coefficients (please see Appendix \ref{App2}). The Poincare sections (see Figures \ref{Schfig1} and \ref{Kerrfig1}) again show the similar behaviour. Hence appearance of the chaotic nature of the particle motion in presence of horizon is an inevitable fact.   

\begin{figure}[htb!]
\centering
\includegraphics[width=7.0cm, height=5.0cm]{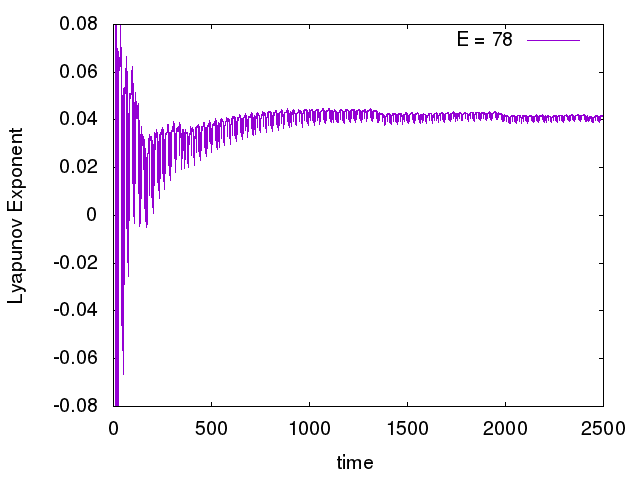}
\caption{(Color online) Largest Lyapunov exponent for the SSS black hole at the energy value $E=78$. The exponent settles at positive value $\sim 0.04$.} 
\label{fig:lypsss}
\end{figure}

In the previous section we analytically predicted that the Lyapunov exponent ($\lambda_L$) has an upper bound given by the surface gravity ($\kappa$). In order to justify our proposition here we present the numerically computed Lyapunov exponents for different cases. We have adopted the standard algorithm to compute the largest Lyapunov exponent which is related to the rate of separation of the trajectories for two nearby points~\cite{sandri:96}.  We consider the energy $E=78$ for the SSS type black hole for which we obtain the full chaotic behaviour of the particles motion. Figure~\ref{fig:lypsss} shows the time variation of the largest Lyapunov coefficient at energy $E=78$. The Lyapunov exponent settles to a positive value ($\sim 0.04$) which supports the chaotic nature of the particle motion. In addition the value is well below the upper bound $\kappa $($=0.25$). 

\begin{figure}[!ht]
\includegraphics[width=7.0cm, height=5.0cm]{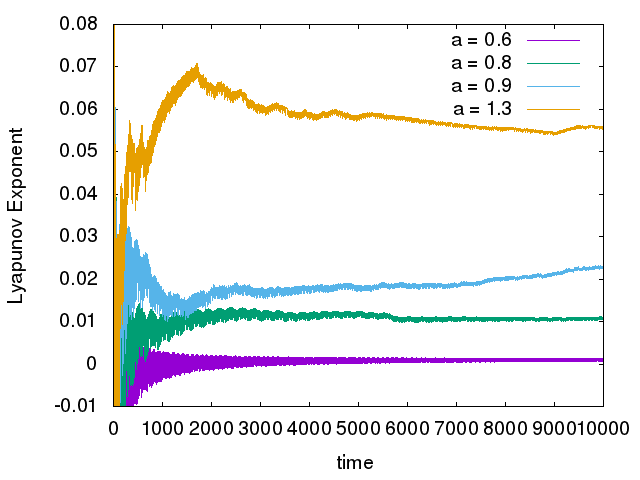} 
\caption{(Color online) Largest Lyapunov exponents for the Kerr black hole for different values of the rotation parameter $a=$ $0.6$, $0.8$, $0.9$ and $1.3$ at constant energy $E=50$. The exponents increases on increase of $a$. } 
\label{fig:lypkerr}
\end{figure}

 In case of Kerr black hole the upper bound on the Lyapunov exponent depends on the rotation parameter (see Eq.~\ref{kerrK}). In Fig.~\ref{fig:lypkerr} we show the time evolution of the largest Lyapunov exponent corresponding to the particle trajectory for Kerr black hole for different rotation parameter ($a=0.6, 0.8, 0.9, 1.3$). Interestingly we observe that the steady state Lyapunov exponent increases on increase of the rotation parameter that supports our observation that the rotation of the black hole induces more chaotic fluctuations in the motion of the particles. In addition we find that the obtained value of the Lyapunov exponents for different values of the $a$ are much lower than the corresponding upper bound (see Eq.~\ref{kerrK}). 

\section{\label{Con}Discussions}
The trajectories of a massless and chargeless particle in a very near horizon region have been studied. The theoretical analysis showed that the radial motion is exponentially increasing function of time parameter. This indicates that the horizon may influence chaotic behaviour in the motion of a particle in an integrable system when it interacts with the horizon of a black hole. Carefully investigating the situation in presence of harmonic potential, it has been observed that this is indeed the case. We found that the trajectories of the particle trapped in a Harmonic potential which is an integrable system,  becomes chaotic in nature after a certain value of energy of the system. Here both SSS and Kerr black holes cases have been investigated. For SSS case the chaos occurs at the particular energy range. Note that our SSS metric is not necessarily restricted to Schwarzschild spacetime, rather it incorporates all candidates of this type. Therefore even the SSS analysis is much more general than the earlier ones. Interestingly, we observed that the rotation parameter induces more chaos to the particle motion. 

Overall we find that although dynamics of a particle is integrable in presence of SSS and Kerr black hole, the addition of harmonic perturbation leads generation of chaos in the system. This is consistent with the KAM theory which states that the nonlinear perturbation in the integrable dynamical system results to the generation of the chaos~\cite{nonlinear:02}. There are several examples reported earlier in support of this claim; like addition of non-linear perturbation in Henon-Heiles potential leads to chaotic behaviour of the system at high energy \cite{ref1}, appearance of chaos in double pendulum (Hamiltonian) for large oscillation \cite{ref2} (also see \cite{ref3} for a discussion in this direction). In our case, for small perturbation, i.e., when the harmonic term is small we observe a regular tori. These tori break and convert into the scattered points as the energy of the system increases which indicate increase in the harmonic perturbation makes the system chaotic. Similar way one can treat the presence of the black hole like a perturbation for the particle trapped in the harmonic potential. Again the integrability of the harmonic particle is lost due to the presence of the black hole. The feature of the non-integrability in the dynamics of the particle that arises due to the addition of the harmonic potential as the perturbation to the particle energy in the presence of the black hole has been reported recently in~\cite{Hashimoto:2016dfz}. This was done for a massive particle. Here we investigated for the massless one.

In this calculation, the outgoing path is taken to be the geodesics for a zero mass particle for which the radial direction is the null geodesic. This particular path plays important role in investigating the Hawking radiation as tunnelling phenomenon \cite{Parikh:1999mf,Banerjee:2008gc}. It reveals that horizon not only radiates quantum mechanically, but also can influence the trajectories of the particles when they are very near to the horizon where the perturbation can come from the other objects in the Universe. If this is the case, then it may be possible that the radiated particles may follow chaotic motion after being escaped from the horizon barrier. Note that due to complexity of the problem here we have not been able to provide any rigorous analytical calculations to support the claim.  To be more sure about the detailed nature of the effect of the horizon on the chaotic fluctuation we need to investigate more on this. Also it must be mentioned that our analysis is completely classical in nature. Probably a systematic calculation based on the quantum mechanics may unravel more information about the chaotic behavior of the particles in black hole spacetime.

Before concluding we also mention that in this paper the influence of horizon on an integrable system has been investigated. Here for simplicity, the integrable system has been chosen as a massless and chargeless particle trapped in harmonic potentials. Then we allowed it to interact with the horizon of a black holes. The underlying idea is exactly in the spirit of the foundations of equilibrium statistical mechanics. In this subject, we always allow our system, which is under investigation, to interact with other system (a finite system for microcanonical ensemble and a heat reservoir for canonical ensemble), either thermally or mechanically. In general it has been assumed that the interaction Hamiltonian is very small (i.e. a weak interaction between the two systems) and hence the total energy of the composite system is sum of the energies of the individual systems. At the equilibrium the question is asked as: What is the probability of finding the system, under investigation, having energy between $E$ to $E+\delta E$ for Microcanonical ensemble while for a canonical ensemble what is the probability of finding the system in a particular microstate? We all know that these information actually encodes the full macroscopic description of the system in terms of number of accessible microstates in Microcanonical ensemble or partition function in Canonical ensemble. In this paper we are also investigating in the same fashion. We allowed one integrable system to interact with the black hole horizon and investigated the trajectories of a particle in the system. The energy of the composite system is taken as the sum of the energies of the individual system, ignoring the interaction; i.e. any influence of the harmonic potentials on the spacetime has not been taken into account. Within this scenario, we found that the composite influence can make particle motion chaotic in nature. Of course, if the interaction is not negligible, then this also has to be taken into account. We leave this for future.


\begin{widetext}
\appendix

\section{{\label{App}}Trajectories in Kerr spacetime}
In the Painleve coordinate transformation the Kerr metric can be written in this form \cite{Jiang:2005ba}:
\begin{eqnarray}
&&d\hat{s}^2=-f dt^2+g dr^2 +2h dtdr +kd\theta^2
\end{eqnarray}
where,
\begin{eqnarray}
&&f=\frac{\bigtriangleup\Sigma}{(r^2 + a^2)^2 - \Delta a^2 \sin^2\theta}~;
\\&&g=\frac{\Sigma}{r^2 + a^2}~;
\\&&h=\frac{\sqrt{2Mr(r^2+a^2)}\Sigma}{(r^2+a^2)^2 - \Delta a^2 \sin^2\theta}~;
\\&&k=\Sigma=  r^2 +a^2 \cos^2\theta~,
\end{eqnarray}
and $\Delta = r^2 +a^2 - 2Mr$.  $M$ is the mass and $a=J/M$ is the angular momentum per unit mass. We call it as rotation parameter. The event horizon is given by $\Delta=0$. It leads to the location of horizon as
\begin{eqnarray}
r_H=M+\sqrt{M^2-a^2}~.
\end{eqnarray}

Using $p_ap^a=0$ and identifying the energy of the massless particle as $E=-p_t$, we find the energy of the particle which is moving through the two harmonic potentials in the above background can be calculated as:
\begin{eqnarray}
E=-\frac{h}{g}p_r +  \sqrt{\frac{h^2}{g^2} p_r^2 +\frac{f}{g} p_r^2 + (\frac{1}{k} \frac{fg+h^2}{g}) p_\theta^2}
+\frac{1}{2}K_r(r-r_c)^2 +\frac{1}{2}K_\theta(y-y_c)^2~.
\label{Eraw}
\end{eqnarray}
Now since we are interested near to the horizon, expanding $f(r)$ upto the first order one obtains
\begin{eqnarray}
f(r)=f(r_H)+(r-r_H)f'(r_H) = -\frac{a^2 \sin^2\theta}{r_H^2 + a^2 \cos^2\theta}+\frac{r_H^4 - a^4 \cos^2\theta + r_H^2 a^2 \sin^2\theta}{
 r_H (r_H^2 + a^2 \cos^2\theta)^2} (r-r_H)~.
\end{eqnarray} 
Using this and replacing all the variables in Eq. (\ref{Eraw}) the energy turns out to be,
\begin{eqnarray}
E=X-\frac{p_rY(a^2+r^2)}{2\Sigma}+\frac{1}{2} \left(y -y _c\right){}^2
   K_{\theta }+\frac{1}{2} \left(r-r_c\right){}^2 K_r~,
\label{E}
\end{eqnarray}
where 
\begin{eqnarray}
&&X=\sqrt{\frac{p_\theta^2+p_r^2(a^2+r^2)}{\Sigma}}~;
\nonumber
\\
&&Y=\sqrt{\frac{(a^2+r_H^2)(a^2+2r^2+a^2\cos 2\theta)(a^2r-2(r-2r_H)r_H^2+a^2r\cos 2\theta)}{r_H(a^2+r^2)(r_H^2+a^2\cos^2\theta)^2}}~.
\end{eqnarray}

Now the equation of motions are
\begin{equation}
\dot{r}=-\frac{(a^2+r^2)(XY-2p_r)}{2X\Sigma}~,
\label{1.33}
\end{equation}
\begin{eqnarray}
&&\dot{p_r}=-K_r \left(r-r_c\right)-\Big[-\frac{r(2X\Sigma-p_rY(a^2+r^2))}{\Sigma^2}+\frac{1}{4Y\Sigma}\{4rX+\frac{2rp_r^2-2rX^2}{X}-2rYp_r
\nonumber
\\
&&-p_r(a^2+r^2)(\frac{Y^2\xi}{r\xi+4r_H^3}+\frac{4rY^2}{a^2+2r^2+a^2\cos 2\theta}-2rY^2) \}\Big]~,
\label{1.34}
\end{eqnarray}
\begin{eqnarray}
\dot{p_\theta}=-\Big[\frac{a^2X\sin 2\theta}{2\Sigma}-\frac{a^2Yp_r(a^2+r^2)\sin 2\theta}{2\Sigma^2}-\frac{Z}{2r_H\Sigma Y\zeta^3}\Big]
-\left(y -y_c\right) r_HK_{\theta }
 \label{1.35}
\end{eqnarray}
and
\begin{eqnarray}
\dot{\theta}=\frac{p_\theta}{X\Sigma}~,
\label{1.36}
\end{eqnarray}
where
\begin{eqnarray}
&&Z=a^2p_r^2(r-r_H)^2(a^2+r_H^2)\Big(a^2(r+2r_H)(1+\cos 2\theta)-2r_H^2(3r+2r_H)\Big)\sin 2\theta~;
\nonumber
\\
&&\zeta=r_H^2+a^2\cos^2\theta~;
\nonumber
\\
&&\xi=a^2-2r_H^2+a^2\cos 2\theta~.
\end{eqnarray}
These equations have been used to study the motion of the particle numerically.

\section{{\label{App2}}Presence of chaos without near horizon approximation of metric coefficients}
Here, for clarity, we present the appearance of chaos without the near horizon approximation of the metric coefficients. For simplicity, a specific SSS has been chosen, which is the Schwarzschild black hole. The Kerr case is also being discussed. This analysis confirms the fact that the horizon in the spacetime indeed induces chaotic behaviour in particle dynamics which is not an artefact of the use of truncated form (i.e. near horizon form of metric coefficient), as we did in our main analysis.
\subsection{Schwarzschild black hole}
For the case of Schwarschild  spacetime the energy of the particle is given by (\ref{1.14}) and the equations of motions are (\ref{1.15}) -- (\ref{1.18}) with $f(r)=1-2M/r$. Here $M$ is the mass of the black hole. Similar to earlier, we again show the Poincare section in ($p_r, r$) plane.
\begin{figure}[!ht]
 \centering
 \includegraphics[scale=0.25, angle=0]{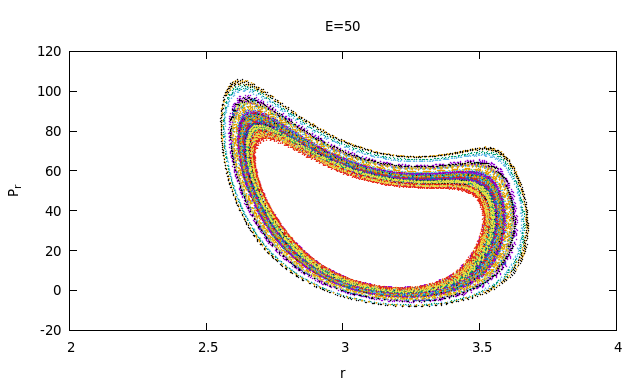}
 \includegraphics[scale=0.25, angle=0]{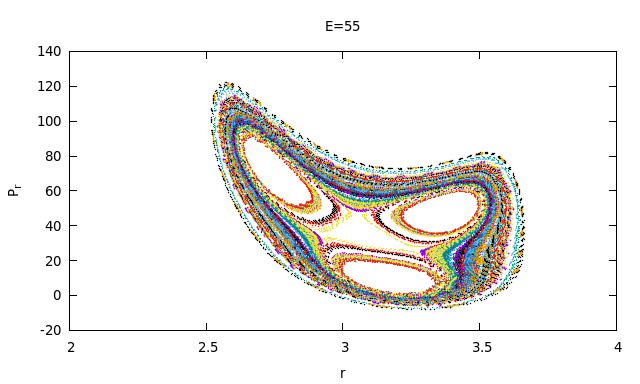}
 \includegraphics[scale=0.25, angle=0]{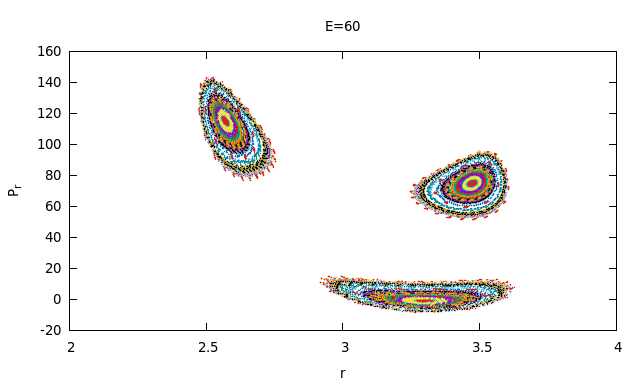}
 \includegraphics[scale=0.25, angle=0]{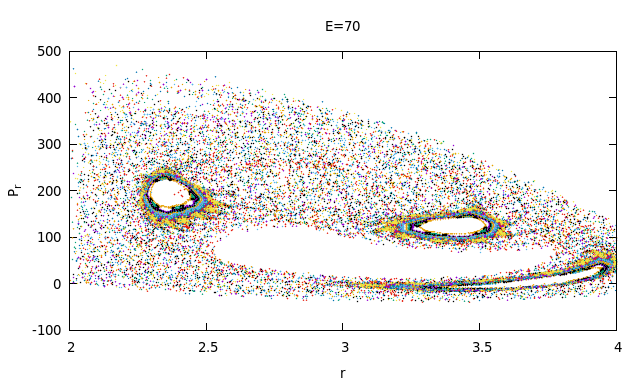}
 \includegraphics[scale=0.25, angle=0]{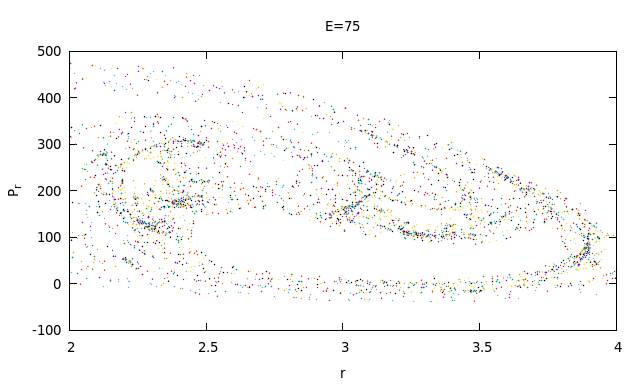}
 \caption{(Color online) The Poincar$\acute{e}$ sections in the ($r$, $p_r$) plane with $\theta=0$ and $p_{\theta}>0$ at different energies for the Schwarzschild black hole. The energies are  $E=50$,  $E= 55$, $E=60$, $E= 70$, and $E= 75$. The other parameters are $r_H=2.0$, $M=1.0$, $r_c=3.2$, $\theta_c=0$, $K_r=100$, and $K_{\theta}=25$.  For large energy the KAM Tori break and the entire region gets filled with the scattered points indicating the presence of chaos.}
 \label{Schfig1}
\end{figure}
\noindent
Figure \ref{Schfig1} clearly shows the presence of chaos in the system after a particular value of energy.

\subsection{Kerr black hole}
Below in Fig \ref{Kerrfig1} we show the appearance of chaos in the system in the case of Kerr black hole without the near horizon approximation.
\begin{figure}[!h]
 \centering
 \includegraphics[scale=0.25, angle=0]{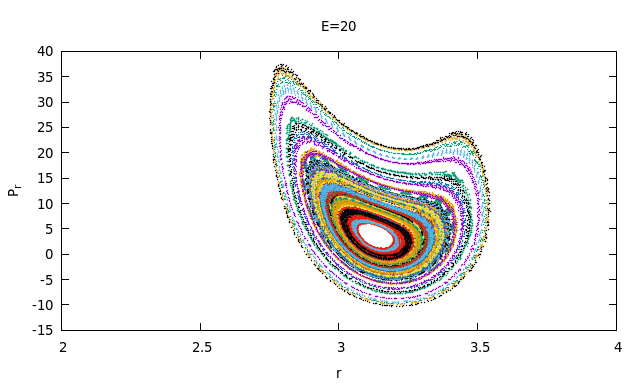}
 \includegraphics[scale=0.25, angle=0]{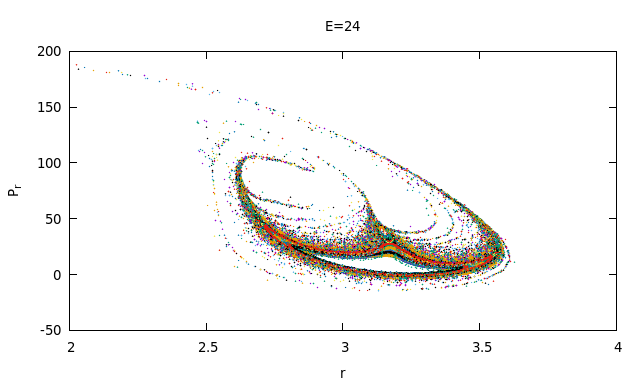}
  \includegraphics[scale=0.25, angle=0]{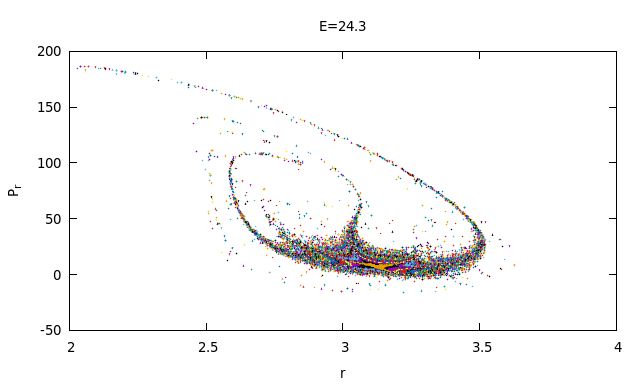}
 \caption{(Color online) The Poincar$\acute{e}$ sections in the ($r$, $p_r$) plane with $\theta=0$ and $p_{\theta}>0$ for different energy for the Kerr black hole model at fixed rotation parameter $a=0.7$. The energies are  $E=20$,  $E= 24$ and $E= 24.3$. The other parameters are same as in Fig.~\ref{Schfig1}.  For large energy the KAM Tori break and the entire region is filled with the scattered points indicating the chaotic trajectory of the particles. }
 \label{Kerrfig1}
\end{figure}
\end{widetext}

\pagebreak


\begin{thebibliography}{99}
\bibitem{comment1} 
 As we are considering the null paths, notion of ``time'' parameter is not so obvious like the case of timelike geodesics. 
Proper time can not be used as a valid parameter since it vanishes along the null path. In this case the four momentum $p^a = dx^a/d\lambda$ is chosen in such a way that the parameter $\lambda$ turns out to be affine one (see analysis around Eq. (3.62) of \cite{Carroll}). The momentum components, appeared in Eq. (\ref{1.04}), are defined in this way. Since $p^a$ is a tensor, the parameter $\lambda$ needs to be a scalar and hence it does not change under coordinate transformation. So the Lyapunov exponent defined here does not has any coordinate dependence.
\end{thebibliography}
\end{document}